\documentstyle[prd,aps,eqsecnum,epsf,psfig,epsfig]{revtex}

\begin{document}
\tighten

\def\bx{\bar x}
\def\by{\bar y}
\def\bz{\bar z}
\def\bt{\bar \theta}
\def\bph{\bar \phi}
\def\bm{\bar m}
\def\bp{\bar p}

\date{today}

\title{Angular Resolution of the LISA Gravitational Wave Detector}
\author{Curt Cutler}
\address{Department of Physics, Pennsylvania State University}
\address{and}
\address{The Albert Einstein Institute, Potsdam}

\date{\today}
\maketitle

\widetext

\begin{abstract}
\hfil
\parbox{5.6in}{We calculate the angular 
resolution of the planned LISA detector, a
space-based laser interferometer for measuring low-frequency 
gravitational waves from galactic and extragalactic sources.
LISA is not a pointed instrument; it is an all-sky monitor with a 
quadrupolar beam pattern.  LISA will measure simultaneously both 
polarization components of incoming gravitational waves, so the data
will consist of two time series. All physical properties of the source, 
including its position, must be extracted from these time series.  
LISA's angular resolution is therefore not a fixed quantity, 
but rather depends on the type of signal and on how much other
information must be extracted.
Information about the source position will be encoded in the 
measured signal in three ways: 1) through the relative amplitudes and phases 
of the two polarization components, 2) through the 
periodic Doppler shift imposed on the signal by the detector's
motion around the Sun, and 3) through the further modulation of the 
signal caused by the detector's time-varying orientation.
We derive the basic formulae required to calculate the LISA's angular 
resolution $\Delta \Omega_S$ for a given source. We then evaluate 
$\Delta \Omega_S$ for two sources of particular interest: 
monchromatic sources and 
mergers of supermassive black holes.  
For these two types of sources,
we calculate (in the high signal-to-noise approximation) 
the full variance-covariance matrix, which gives the 
accuracy to which all source parameters can be measured.
Since our results on LISA's angular resolution depend mainly on 
gross features of the detector geometry, orbit, and noise curve, 
we expect these results to be fairly insensitive to modest changes in 
detector design that may occur between now and launch.  
We also expect that our calculations
could be easily modified to apply to a modified design.}\hfil

\end{abstract}


\twocolumn
\narrowtext
\section{Introduction}\label{intro}

This paper calculates the angular resolution of
the planned LISA gravitational wave detector.  
LISA (short for Laser Interferometer Space Antenna) is in essence
a space-based version of the ground-based interferometric detectors
currently under construction: LIGO, VIRGO, etc. 
There are some major differences, however. The primary difference is that
LISA will be sensitive to gravitational waves in a much 
lower frequency band:
$10^{-4}-10^{-1} {\rm Hz}$.  
(This low-frequency regime is unobservable by any
proposed ground-based detectors, due to seismic noise.
The ground-based interferometers will be sensitive 
in the range $10^1-10^3 \,{\rm Hz}$.) 
The $10^{-4}-10^{-1} \,{\rm Hz}$ band contains many known gravitational 
waves sources that LISA is `guaranteed' to see.  These guaranteed sources
comprise a wide variety of short-period binary star systems, both
galactic and extragalactic, including: close white dwarf binaries, 
interacting white dwarf binaries, unevolved binaries, W UMa binaries,
neutron star binaries, etc.~\cite{lp:87,lpp:87,Hils:1990}. 
Indeed, our galaxy probably contains so many short-period, stellar-mass 
binaries that
LISA will be unable to resolve them individually, and
the resulting `confusion noise' will actually dominate over instrumental
noise as the principal obstruction to finding {\it other} sources of 
gravitational waves in the datastream.
Besides stellar-mass binaries, other possible LISA sources 
include: 
1) a stochastic GW background generated in the early universe, 
2) the inspiral of compact, stellar-mass objects into 
supermassive black holes (SMBH's), 
and 3) the merger of two SMBH's~\cite{Pre}. 
The detection of any one of these would 
clearly be of immense interest.   
The events involving supermassive black holes must surely 
occur in the universe, but the event 
rates are highly uncertain.

Besides their different frequency bands, another important difference 
between LISA and the ground-based
interferometers concerns their means of identifying the angular 
position of the source on
the sky. LISA is not a pointed instrument; it is an `all-sky monitor' with 
a quadrupolar beam pattern.   The ground-based detectors share this 
characteristic, but because there will be at least three ground-based
detectors, and because they will be sensitive to gravitational radiation 
whose wavelength is much shorter than the distance between detectors, 
they will be able to determine the source position to within
$\sim 1^o$ by a standard time-of-flight method.  This method is not
available to LISA. Only one
space-based detector is currently planned. Moreover the gravitational 
wavelength at the heart of the LISA band ($\sim 10^{-3}$ Hz) is 
of order 1 AU, so a second detector would have to be placed at least 
several AU away from the Earth for time-of-flight 
measurements to give useful constraints on source positions.

As we shall see, LISA can be thought of as two detectors, each measuring
a different polarization of the gravitational wave. Thus the data consists of two
time series. All information about
the source position (as well as {\it all other} physical variables) must be 
extracted from these two time series.
Angular position information is encoded in the time series in the following ways.
First, the relative amplitudes and
phases of the two polarizations provide some position information.
Second, most sources will be `visible' to LISA for months or
longer, so LISA's 
translational motion  around the Sun
imposes on the signal a  periodic Doppler shift, whose magnitude and 
phase depend on the angular position of the source. 
(In exactly the same way, radio astronomers take advantage 
of the Doppler shift caused by the Earth's rotation to determine a
pulsar's position to an accuracy much better than that implied 
by the beam width of the radio telescope.)
Finally, as we shall describe, LISA's orientation rotates
on a one-year timescale, which imposes a further source-position-dependent
modulation on the measured signal.
In Fourier space, the effect of the detector's changing orientation on
a monochromatic signal of frequency $f_0$ is to spread 
the measured power over (roughly) a range $f_0 \pm 2/T$, where T is one year.
(The factor of $2$ arises from the quadrupole beam pattern of the
detector.) The effect of the periodic Doppler shift coming from 
the detector's center-of-mass motion is to spread the power over a
range  $f_0(1 \pm v/c)$, where $v/c \sim 10^{-4}$.  
These two effects are therefore of roughly
equal size at $f_0 \sim 10^{-3} \,{\rm Hz}$, which is near the center of
the LISA band; rotational modulation is more significant at lower frequencies 
and Doppler modulation is more significant at higher frequencies.

It is also worth emphasizing that much of the 
uncertainty in the position measurement 
arises from the fact that from this pair of time series one must try to 
extract {\it all} the physical parameters of the binary: 
the orbital plane of the binary,
the masses of the bodies, etc.
Errors in determining the source position are correlated with errors in 
these other parameters.  The result is that 
LISA's angular resolution is significantly 
worse than one would suppose if one ignored these correlations.
From this consideration, it should be clear that LISA's angular resolution depends 
not just on the detector and the signal-to-noise, but
on the type of source as well.

In this paper we derive the formulae necessary 
for calculating LISA's angular resolution $\Delta \Omega_S$ for some 
given source, and we then evaluate 
$\Delta \Omega_S$ for two sources of particular interest: 
monochromatic sources (e.g., short-period, white-dwarf binaries) and 
mergers of supermassive black holes.  For these two types of sources,
we perforce calculate how accurately LISA can measure all the other
sources parameters, as well. 
A preliminary estimate of LISA's angular resolution 
has already been made by Peterseim et al.~\cite{Pre,Peterseim_etal:1996},
but that estimate was only for
monochromatic, high-frequency sources, and it assumed that the frequency,
polarization, and amplitude of the signal were known a priori, so only the
source position had to be extracted from the data. Also, for simplicity 
the estimate by Peterseim et al.~\cite{Pre,Peterseim_etal:1996} 
took into account the information from only a 
single polarization, and it neglected the information encoded via the 
rotation of the detector.  
In essence, our paper provides a much more realistic calculation. 
Since LISA's angular resolution depends mainly on gross features of
the detector orbit and noise curve, rather than exact 
details of the detector design, 
we expect that our results will be fairly insensitive to contemplated
design changes. We also expect that the calculation presented here could be
very easily modified to apply to a different design.

We now turn to our main motivation for considering SMBH mergers. 
The striking feature of these mergers is the 
huge amplitude of the emitted gravitational waves. 
(That is, huge compared to other GW sources!)
LISA would be capable of detecting SMBH mergers
at basically any reasonable redshift ($z < 10$, say) 
with signal-to-noise $S/N \sim 10^3$, so long as the black holes 
are in the mass range $10^4 M_\odot < M(1+z) < 10^7 M_\odot$.
This mass range is set by the frequency band where LISA is sensitive.
While the event rate for such mergers is highly uncertain 
(it could be several per year, or $<< 1$/yr; see~\cite{vecchio_rate} for
a recent review), 
if SMBH mergers {\it were} discovered, they could provide a 
way of determining all the basic cosmological parameters -- $H_0, \Omega_0$, and $\Lambda_0$ -- to remarkably high accuracy.  
The idea is that from the gravitational waveform
one expects determine the luminosity distance $D_L$ to the source to an accuracy of roughly $(S/N)^{-1}$. (In fact, we find that LISA does 
roughly a factor $10$ worse than this: typically $\Delta D_L/D_L \sim 1\%$; see Table 2).  As pointed out in the LISA Pre-Phase A Report~\cite{Pre} 
(hereinafter  referred to as the LPPAR), if the source position on the sky could 
be determined to sufficient accuracy that one could identify the host galaxy 
or galaxy cluster, then presumably one could also determine the redshift optically.  Clearly, a mere handful of such
measurements would be sufficient to determine 
$H_0, \Omega_0$, and $\Lambda_0$ to roughly $1\%$ accuracy.  
One of the key motivations for this paper is to see whether 
LISA has sufficient angular resolution to make such identifications 
possible.  In brief, we find that LISA will determine the SMBH location to
no better than $\sim 10^{-5}$ steradians (and typically to $\sim 10^{-4}$ 
steradians) which is not sufficient by itself 
to permit identification of the host galaxy.  However this 
position measurement will be 
available days before the final merger; other telescopes (radio, optical,
X-ray) should know when and where to look, so if the 
merger is accompanied by an electromagnetic outburst
(e.g., from an accretion disk being carried along by one of the holes), 
the host galaxy might still be determined. 
Whether such an electromagnetic outburst
can be expected appears to be an interesting 
open problem \cite{bbr}.

The plan of this paper is as follows.  Secs.~II and III
give brief summaries of relevant background information.  
Sec.~II reviews the basic formulae
of signal analysis and parameter estimation (mostly 
to establish notation and conventions), while 
Sec.~III describes LISA's configuration and orbit, 
its response to gravitational waves, and its noise sources. 
In Sec.~IV we derive LISA's angular resolution for monochromatic sources,  
and in Sec.~V we derive LISA's angular resolution for SMBH mergers.
For both cases we calculate the Fisher
matrix, which estimates how accurately the detector can extract 
{\it all} the physical parameters of the system from the measured signal. 
Our conclusions are summarized in Sec.~VI. 

We should state at the outset the principal limitations of this study.
Firstly, and necessarily, the current detector design cannot be
regarded as final, and the published noise curves--for both the
instrumental noise and the confusion noise--can only be regarded as
rough estimates. {\it Faute de mieux}, we use the current design 
and the best estimates of the noise sources currently available.
Relatedly, while it is a goal of the LISA design that the instrumental noise 
should be stationary and Gaussian\cite{Pre}, probably we will not 
know how well this goal has been met until the 
the instrument is functioning.  {\it Faute de mieux}, we assume the
noise is stationary and Gaussian.
Finally, to simplify things, in our treatment of SMBH mergers
we assume that the two holes are in a circular orbit, and we
assume that the plane of the binary orbit is fixed. That is, 
we ignore both orbital eccentricity and the precession of
the orbital plane caused by the spins of the holes \cite{haris}.  
We intend to take both these
effects into account in a later paper~\cite{alberto_inprogress,alberto_curt}.

Throughout this paper, we assume that the detector arm length is
much smaller than the gravitational wavelength, and
we take the gravitational waveform $h_{ab}(t)$ to be in the 
(standard) de Donder gauge. Time is in units of seconds 
and frequency is units of Hz.

\section{Review of Signal Analysis}

This section briefly reviews the basic formulae of signal
analysis, partly to fix notation.  For a more complete
discussion, see  \cite{cutler_flanagan} or \cite{WZ}. 

The output of $N$ detectors can be represented by the
vector $s_\alpha(t)$, $\alpha = 1,2,...,N$. It is often convenient
to work with the Fourier transform of the signal; the convention we use is
\begin{equation}
\label{fourierT}
{\tilde s_\alpha}(f) \equiv \int_{-\infty}^{\infty}\, e^{2\pi i f t}s_\alpha(t)\, dt,
\end{equation}
 
The output $s_\alpha(t)$ is the sum of gravitational waves
$h_\alpha(t)$ plus instrumental noise $n_\alpha(t)$. 
We assume that the noise is stationary and Gaussian.
`Stationarity' essentially means that the different Fourier components 
$\tilde n_\alpha(f)$ of the noise are uncorrelated; thus we have 
\begin{equation}
\label{pr}
\langle {\tilde n_\alpha}(f) \, {\tilde n_\beta}(f^\prime)^* \rangle = {1 \over 2}
\delta(f - f^\prime) S_n(f)_{\alpha \beta},
\end{equation}
where `$\langle \ \rangle$' denotes the `expectation value' and
$S_n(f)_{\alpha \beta}$ is referred to as the spectral 
density of the noise. When N=1 (i.e., when there is just a single
detector), we will dispense with Greek indices and 
just write $\tilde s(f)$ and $S_n(f)$.
 
`Gaussianity' means that each Fourier component has 
Gaussian probability distribution.  
Under the assumptions of stationarity and Gaussianity, we obtain a 
natural inner product on the vector space of signals.  
Given two signals $g_{\alpha}(t)$
and $k_\alpha(t)$, we define $\left( {\bf g} \, | \, {\bf k} \right)$ by
\begin{equation}
\label{inner}
\left( {\bf g} \,|\, {\bf k} \right) = 2 \int_0^{\infty} \,[S_n(f)^{-1}]^{\alpha \beta}
\biggl( \tilde g_\alpha^*(f)
\tilde k_\beta(f) + \tilde g_\alpha(f) \tilde k_\beta^*(f)\biggr) \,  df
\end{equation}
\noindent In terms of this inner product, the probability for the 
noise to have some realization
${\bf n}_0$ is just
\begin{equation}
\label{pn0}
p({\bf n} = {\bf n}_0) \, \propto \, e^{- \left( {\bf n}_0\, |\, {\bf n}_0 
\right) /2}.
\end{equation}
Thus, if the actual incident waveform is ${\bf h}$, the probability of measuring a signal ${\bf s}$ in the
detector output is proportional to 
$e^{-\left( {\bf s-h} \, | \, {\bf s-h}\right)/2}$.  Correspondingly, given a measured signal ${\bf s}$, the
gravitational waveform ${\bf h}$ that ``best fits'' the data is the one that
minimizes the quantity $\left( {\bf s-h} \, | \, {\bf s-h} \right)$.

It also follows from Eq.~(\ref{inner}) that for any functions $g_\alpha(t)$
and $k_\alpha(t)$,
the expectation value of $({\bf g}|{\bf n}) ({\bf k}|{\bf n})$,  
for an ensemble of realizations of the 
detector noise $n_\alpha(t)$, is just $({\bf g}|{\bf k})$.
Hence the signal-to-noise of the detection will be
approximately given by 
\begin{equation}
\label{snh}
{S\over N}[ {\bf h}] = 
{{({\bf h}|{\bf h})}\over { {\rm rms}\  ({\bf h}|{\bf n})}} = 
({\bf h}|{\bf h})^{1/2}.
\end{equation}

For a given incident gravitational wave, different realizations
of the noise will give rise to somewhat different best-fit
parameters.  However, for large $S/N$, the best-fit parameters will have a
Gaussian distribution centered on the correct values. 
Specifically, let ${\tilde \lambda}^i$ be the ``true'' values of the
physical parameters,
and let ${\tilde \lambda}^i + \Delta \lambda^i$ be the best
fit parameters in the presence of some realization of the noise.  Then
for large $S/N$, the parameter-estimation errors $\Delta \lambda^i$ have
the Gaussian probability distribution
\begin{equation}
\label{gauss}
p(\Delta \lambda^i)=\,{\cal N} \, e^{-{1\over 2}\Gamma_{ij}\Delta \lambda^i
\Delta \lambda^j}.
\end{equation}
Here $\Gamma_{ij}$ is the so-called Fisher 
information matrix  defined
by
\begin{equation}
\label{sig}
\Gamma_{ij} \equiv \bigg( {\partial {\bf h} \over \partial \lambda^i}\, \bigg| \, 
{\partial {\bf h} \over \partial \lambda^j }\bigg),
\end{equation} 
and ${\cal N} = \sqrt{ {\rm det}({\bf \Gamma} / 2 \pi) }$ is the
appropriate normalization factor.  For large $S/N$, the 
variance-covariance matrix is given by
\begin{equation}
\label{bardx}
\left< {\Delta \lambda^i} {\Delta \lambda^j}
 \right>  = (\Gamma^{-1})^{ij} + {\cal O}(S/N)^{-1} \;.
\end{equation}

\section{LISA}

In this section we describe LISA's geometry and noise curve; 
these are the only aspects of the detector that are necessary
for our analysis. For more details on how the detector works, 
we refer the reader to the LPPAR~\cite{Pre}.

\subsection{Detector Configuration and Orbit}

The geometry of the LISA mission, as currently envisioned, is depicted in 
Figs.~1 and 2. The detector is a three-arm laser interferometer, with
each arm being $5 \times 10^6$ km long. It consists of six drag-free satellites,
positioned so that two adjacent satellites sit at each vertex of an
equilateral triangle.
(One satellite at each vertex would suffice, but the current design
calls for two, which provides some redundancy and greatly simplifies 
the pointing control.)
The detector's center-of-mass follows a circular, heliocentric trajectory, 
trailing $20^o$ behind the Earth.  
The plane of the detector is tilted by $60^o$
with respect to the ecliptic; this angle allows the 
satellites to {\it maintain} the shape of an equilateral triangle 
throughout the orbit.  We refer the reader to Fig. 3.17 (p.92) of 
the LPPAR
for a useful picture of the orbital geometry, and to Faller et al.~\cite{faller_etal}
for a simple explanation of {\it why} the $60^o$-tilt allows the
equilateral shape to be maintained.

We label the arms $1,2,3$, and call their lengths $L_1,L_2,L_3$.
Gravitational waves cause time-varying changes in arm lengths 
by amounts $\delta L_i(t)$, which are different in the three arms.
The differences are measured interferometrically \cite{foot2}.  
With 3 arms, there are two 
linearly independent differences: $\delta L_1 - \delta L_2$ and 
$\delta L_2 - \delta L_3$.  
Therefore LISA will be able to measure
simultaneously both polarizations of an incoming gravitational wave. 

We find it useful to introduce {\it two} Cartesian coordinate systems: 
`unbarred' coordinates $(x,y,z)$ tied to the detector and 
`barred'  coordinates $(\bx,\by,\bz)$ tied to the ecliptic. 
Unbarred and barred spherical polar angles $(\theta, \phi)$ and $(\bt, \bph)$
are related in the usual way to the Cartesian coordinates: 
cos$\,\theta = z/(x^2+y^2+z^2)^{1/2}$,  
cos$\,\bt = \bz/(\bx^2+\by^2+\bz^2)^{1/2}$, etc.  
We denote by $x^a$ the unit vector along the x-axis, and similarly
for $y^a, z^a, \bx^a, \by^a, \bz^a$. Here the superscript `a' is an abstract index indicating that the object is a vector in 3-dimensional space. 
The unit vectors along the arms are called $l_1^a,l_2^a,l_3^a$, respectively.  
The detector lies in the $x-y$ plane, and the $x-y$ coordinates rotate with 
detector.  
We assign, for all time, the following
coordinate-directions to the three arms:
\begin{equation}
l_i^a = {\rm cos}\gamma_i \, x^a + {\rm sin}\gamma_i \, y^a
\label{a1}
\end{equation}
\noindent where
\begin{equation}
\gamma_i = \pi/12 \ + \ (i-1)\pi/3 
\label{a2}
\end{equation}

\noindent as shown in Fig.~2. The $\pi/12$ term in Eq.~(\ref{a2}) is included
for later convenience. 
We choose the $\bx-\by$ plane to be the ecliptic (i.e., the plane
of the Earth's motion around the Sun).
The detector's center-of-mass follows the trajectory
\begin{equation}
\bt(t) = \pi/2, \ \ \ \ \bph(t) = \bph_0 + 2\pi t/T
\label{cmt}
\end{equation}
\noindent where $T$ equals one year, and where $\bph_0$ is just a constant
that specifies the detector's location at time $t=0$.

The normal to the detector plane, $z^a$, is at a constant $60^o$ angle
to $\bz^a$, and $z^a$ precesses around $\bz^a$ at a constant rate:
\begin{equation}
z^a = \case12 \bz^a - \case{\sqrt{3}}2 
\biggl({\rm cos}\bph(t)\, \bx^a 
+ {\rm sin}\, \bph(t) \by^a \biggr)\;.
\label{za}
\end{equation}
Using $z_a l_i^a = 0$ and Eqs.~(\ref{a1})--(\ref{za}), we see
that the directions $l_i^a$ can be written in terms of the 
barred coordinates as
\begin{eqnarray}
\label{lbar}
l_i^a &=& {\rm cos}\,\alpha_i(t) \,\biggl[{\rm cos}\,\bph(t) \, \by^a - 
{\rm sin}\,\bph(t) \, \bx^a\biggr] \nonumber \\
&+& \, {\rm sin}\, \alpha_i(t)\,\biggl[\case{\sqrt{3}}2 \, \bz^a + 
\case12\bigl({\rm cos}\,\bph(t) \,\bx^a 
+ {\rm sin}\,\bph(t) \,\by^a\bigr) \biggr], 
\end{eqnarray}
where the $\alpha_i(t)$ increase linearly with time:
\begin{equation}
\alpha_i(t) = 2\pi t/T - \pi/12 - (i-1)\pi/3 + \alpha_0 \;,  
\label{alphai}
\end{equation} 
where $\alpha_0$ is just a constant specifying the orientation of
the arms at $t=0$.

In this paper we are primarily interested in LISA's angular resolution.
The error box for the position 
measurement has solid angle $\Delta \Omega_S$ given by
\begin{equation}
\Delta \Omega_S  = 2 \pi\,\biggl[\Delta {\bar\mu}_S \, \Delta {\bar\phi}_S 
- \left< \Delta {\bar\mu}_S \, \Delta {\bar\phi}_S \right> \,\biggr]
\label{solid}
\end{equation}
\noindent where ${\bar\mu_S} \equiv {\rm cos}\,\bt_S$.
The second term in brackets in Eq.~\ref{solid} accounts for the fact that 
errors in $\mu_S$ and $\phi_S$ will in general be correlated, so
that the error box on the sky is elliptical in general, not circular.
Also note the overall factor of $2\pi$ in our definition of $\Delta \Omega_S$. 
With this choice, 
the probability that the source lies {\it outside} an
(appropriately shaped) error ellipse enclosing solid angle 
$\Delta \Omega$ is 
simply $e^{-\Delta \Omega/\Delta \Omega_S}$.

\begin{figure}[h]
\epsfxsize =210pt \epsfbox{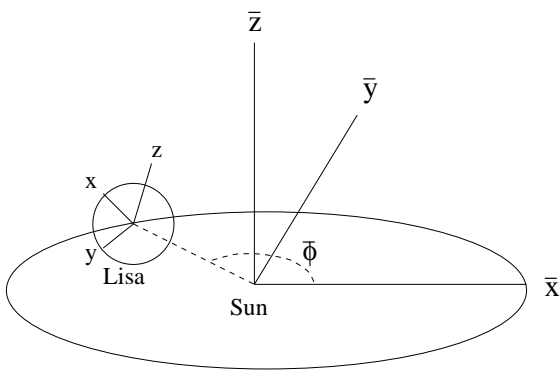}
\vskip 1cm
\caption{\label{figure1} Shows the two coordinate systems used in
our analysis: `barred' coordinates tied to the ecliptic and `unbarred'
coordinates that are tied to the detector and rotate with it.}
\end{figure}

\begin{figure}[h]
\epsfxsize =210pt \epsfbox{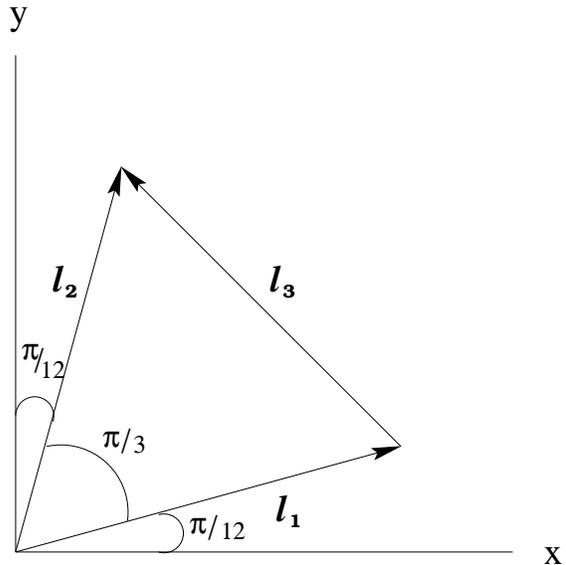}
\vskip 1cm
\caption{\label{figure2} Illustrates the orientation of LISA's 3 arms
in the $x-y$ plane.}
\end{figure}

\subsection{Detector Response}

Because the LISA antenna has three arms, it produces 
basically the same information
as a pair of two-arm detectors, and therefore is
capable of simultaneously measuring both polarizations of the 
gravitational wave.  To begin with, however, we shall consider
only a {\it single} two-arm detector, formed by arms $1$ and $2$.
The extension to two detectors will be straightforward. 

\subsubsection{Single Detector}

We refer to the detector formed just by arms $1$ and $2$ 
as `detector I.'
Detector I measures $h_I(t)$, given by
\begin{mathletters}
\label{hdef}
\begin{eqnarray}
h_I(t) &=& \biggl(\delta L_1(t) - \delta L_2(t)\biggr)/L
\label{hd1}\\
&=& \case12 \, h_{ab}(t)\bigl(l_1^al_1^b - l_2^al_2^b\bigl)
\label{hd2}\\
&=& \case{\sqrt{3}}2 \bigl(\case12 h_{xx} - \case12 h_{yy}\bigr)
\label{hd3}
\end{eqnarray}
\end{mathletters}
\noindent where we used Eq.~(\ref{a1}) to go from~(\ref{hd2}) 
to~(\ref{hd3}). 
That is, for an arbitrary waveform
$h_{ab}(t)$, the `relative arm length difference' $h_I(t)$ measured 
by a $60^o$-interferometer 
is always precisely $\sqrt{3}/2$ times as large as the 
`relative arm length difference' measured by a $90^o$-interferometer 
(a result which is well-known among LISA cognoscenti). 
This result assumes the
$60^o$-interferometer is `placed symmetrically' inside the
$90^o$-interferometer, as shown in Fig.~2. That is the reason for
our $\pi/12$ term in Eq.~(\ref{a2}).
This simple observation saves a lot of work. It means that all the 
formulae derived in the extensive literature on $90^o$-interferometers 
(LIGO, VIRGO, etc.) can be carried over immediately to LISA: one just 
remembers to multiply the signal amplitude by $\sqrt{3}/2$.

Consider a monochromatic, plane-fronted gravitational wave travelling
in the $-n^a$ direction.  The general such solution can be written
as the sum of two orthogonal polarization states. 
Let $p^a$ and $q^a$ be axes 
orthogonal to $n^a$, with $q^a \equiv -\epsilon^{abc} n_b p_c$. Define polarization basis tensors by 
\begin{equation}
H^+_{ab} = p_a p_b - q_a q_b  \ , \ \ H^\times_{ab} = p_a q_b + q_a p_b \;.
\label{poldef}
\end{equation}  
For a particular, unique choice of $(p^a,q^a)$, called
the waves' {\it principal axes}, there is precisely a
$\pi/2$ phase delay between the two polarizations:
\begin{equation}
h_{ab}(t) = A_+ H^+_{ab} \cos(2\pi f t) +  A_\times H^\times_{ab} \sin(2\pi f t)\,
\label{elliptical}
\end{equation}
Here $A_+$ and $A_\times$ are
constants, the amplitudes of the two polarization states,
and we have omitted an arbitrary phase by our choice of the zero of time.
Our convention is $A_+ \ge |A_\times| \ge 0$; $A_\times \ge 0$ for  
right-hand  polarized waves  and $A_\times \le 0$ for left-hand polarization. 

The strain $h_I(t)$ that the waves produce in 
detector I depends on $A_+$ and $A_\times$, the 
principal polarization axes, and the direction of propagation: 
\begin{eqnarray}
\label{ht}
h_I(t) &= \case{\sqrt{3}}2 A_+ F_I^+(\theta_S,\phi_S,\psi_S) \cos(2\pi f t) 
\nonumber \\
\mbox{} &+ \case{\sqrt{3}}2 A_{\times} F_I^{\times}(\theta_S,\phi_S, \psi_S) \sin(2\pi f t)\;,
\end{eqnarray}
where $F_I^+$ and $F_I^\times$ are the ``detector beam-pattern''
coefficients~\cite{300years}:
\begin{mathletters}
\label{Fpc}
\begin{eqnarray}
F_I^+(\theta_S,\phi_S,\psi_S) &= \case12(1 + \cos^2 \theta_S) \cos 2\phi_S \cos
2\psi_S \cr
\mbox{}&-\cos \theta_S \sin 2\phi_S \sin 2\psi_S\;,
\label{Fp}
\\
F_I^{\times}(\theta_S,\phi_S, \psi_S) &= \case12(1 + \cos^2 \theta_S) \cos 
2\phi_S
\sin 2\psi_S \cr
\mbox{}&+ \cos \theta_S \sin 2\phi_S \cos 2\psi_S\;.
\label{Fc}
\end{eqnarray}
\end{mathletters}
\noindent Here the subscript `S' stands for `source',
$(\theta_S,\phi_S)$ give the source location in the `unbarred', 
detector-based coordinate system, and $\psi_S$ 
is the `polarization angle' of the wavefront, defined
(up to an arbitary multiple of $\pi$) as follows:
\begin{equation}
{\rm tan} \,\psi_S = (z^a q_a)/(z^b p_b)
\label{psi_general}
\end{equation}

In the case of interest here, the source
polarization is assumed fixed, but the detector plane rotates throughout
the observation.  A very closely related problem was examined by
Apostolatos et al.~\cite{haris} (hereinafter referred to as ACST), 
who investigated the case 
where the detector location and orientation are fixed 
(on the timescale of the observation) but where the orbital
plane of the binary undergoes Lense-Thirring precession (due
the post-Newtonian coupling between the bodies' spins and their
orbital angular momentum), thereby modulating the 
complex amplitude of the measured signal in a nearly equivalent way.
We refer to ACST for derivations of many of the formulae quoted below.
(There is  one important difference between the `rotating-source' 
case studied in ACST and the `rotating-detector' case studied here:
the `Thomas precession' term identified in ACST is absent in the 
`rotating-detector' case. Note also that some of the sign conventions 
in ACST are different from those used here; in particular, $A_{\times}$
is defined to be positive in ACST, but has no definite sign here.)

To begin, we rewrite the signal (\ref{ht})
in the conventional amplitude-and-phase form. For a waveform 
whose amplitude and frequency are slowly changing functions of time,
we can write
\begin{equation}
h_I(t)= \case{\sqrt{3}}2 \, A_I(t) \cos [\int^t{2\pi f(t') dt'} + \varphi_{p,I}(t) 
+ \varphi_D(t)]\;.
\label{h_amp_phase}
\end{equation}
where $f(t)$ is the GW frequency that {\it would} be measured by a non-rotating
detector positioned at the solar system barycenter, and  
$A_I(t)$, $\varphi_{p,I}(t)$, and $\varphi_D(t)$ are given by
\begin{mathletters}
\label{amp_phase}
\begin{eqnarray}
A_I(t) &=& \biggl(A_+^2 {F_I^+}^2(t)  + A_\times^2 {F_I^\times}^2(t) \biggr)^{1/2} \;,
\label{amp}\\ 
\varphi_{p,I}(t) &=& \tan^{-1}\left({{-A_\times F_I^\times(t)}
\over {A_+ F_I^+(t)}}\right)\;,
\label{pphase}\\
\varphi_D(t) &=& 2\pi f(t) c^{-1}\,  R \, {\rm sin}\,\bt_S \ 
{\rm cos}\biggl(\bph(t) - \bph_S\biggr)  \;.
\label{Dphase}
\end{eqnarray}
\end{mathletters}

\noindent where $R = 1 {\,\rm AU}$. 
We refer to $\varphi_D(t)$ as the {\it Doppler phase};
it is just the difference betweem the phase of the wavefront at the detector
and the phase of the wavefront at the barycenter.  
We have neglected second-order Doppler corrections to Eq.~(\ref{Dphase});
this is justified since such corrections are of order
$(v/c) |\varphi_D(t)| \alt 3 \times 10^{-4} (f/10^{-3})$ radians. 
In~(\ref{Dphase}) we
also neglect the small change in the source frequency $f$ (as measured
at the barycenter) that occurs {\it during} the time delay 
$R\, {\rm sin}\bt_S/c$; the fractional correction to $\varphi_D(t)$ due to
this effect is of order $\case12 f^{-1} (df/dt) R/c$, which for a binary is 
$\sim 0.04 (4\mu/M) (6 M/r)^{5/2} (f/10^{-3})$, where
$\mu$ and $M$ are the reduced and total mass of the binary, respectively,
and $r$ is the orbital radius.

$A_I(t)$ is the waveform amplitude, and we refer to 
$\varphi_{p,I}(t)$ as the waveform's {\it polarization phase}.
The time-dependence of $A_I(t)$ and $\varphi_{p,I}(t)$ are determined
by $F_I^+(t)$ and $F_I^\times(t)$, which in turn depend on
$\theta_S(t),\phi_S(t), {\rm and}\, \psi_S(t)$.
Using Eqs.~(2.1)--(2.6) we find
\begin{equation}
\cos\theta_S(t) = \case12 \cos\bt_S - \case{\sqrt{3}}2\sin\bt_S
\cos(\bph(t)-\bph_S) \;\label{cts} 
\end{equation}

\FL
\begin{eqnarray}
\label{pts}
\phi_S(t) &=& \alpha_1 + \pi/12 \nonumber \\
&-& \tan^{-1}
\biggl[{ { \sqrt{3}\cos\bt_S \ + \ 
\sin\bt_S \cos(\bph(t)-\bph_S)}\over 
{2\sin\bt_S\cos(\bph(t)-\bph_S)} }\biggr]\;
\end{eqnarray}

Now consider the case where the monochromatic source is a circular,
Newtonian binary. (This is quite general: any monochromatic
point source can be represented as a circular, Newtonian binary.)
The lowest-order, quadrupole approximation gives:
\begin{mathletters}
\label{binary}
\begin{eqnarray}
p^a &=& \epsilon^{abc} n_b \hat L_c,
\label{b1}\\ 
A_+ &=& {{2 M_1 M_2}\over {r D}}\biggl[1 + (\hat L^a n_a)^2
\biggr] \;,
\label{b2}\\
A_{\times} &=&   -{{4 M_1 M_2}\over {r D}} \hat L^a n_a  \;.
\label{b3}
\end{eqnarray}
\end{mathletters}

\noindent 
where $M_1$ and $M_2$ are the two masses, $r$ is their orbital
separation, $D$ is the distance between source and observer, and  
$\hat L^a$ is the unit vector parallel to the 
binary's orbital angular momentum vector.
The binary's circular orbit, when projected on the plane of the sky--
i.e., projected orthogonal to the waves' propagation 
direction--looks elliptical. The principal axis $p^a$ is just 
the major axis of this ellipse. (See Fig.~21 in ACST.)
We let $\hat L^a$ point in the direction $(\bt_L,\bph_L)$.  
The angles $\theta_S(t)$ and $\phi_S(t)$ do not depend on 
the principal polarization direction $p^a$, so they are already given
by Eqs.~(\ref{cts})--(\ref{pts}) above. Using Eqs.~(\ref{psi_general}) and (\ref{b1}), 
$\psi_S(t)$ is given by
\FL
\begin{equation}
\tan \psi_S(t) =  \biggl(\hat L^a z_a -   \hat L^a n_a z^b n_b\biggr)\biggl/
\biggl(\epsilon_{abc}n^a \hat L^b z^c\biggr)   
\label{psis}
\end{equation}
where
\FL
\begin{equation}
\label{lz}
\hat L^a z_a = \case12 {\rm cos}\,\bt_L - \case{\sqrt{3}}2 
{\rm sin}\; \bt_L  {\rm cos}\,\bigl(\bph(t) - \bph_L \bigr)
\end{equation}
\FL
\begin{equation}
\label{ln}  
\hat L^a n_a = {\rm cos}\; \bt_L {\rm cos}\;\bt_S +
{\rm sin}\,\bt_L  \; {\rm sin}\,\bt_S \; 
{\rm cos}\,\bigl(\bph_L - \bph_S \bigr) \;
\end{equation}
\FL
\begin{eqnarray}
\label{nlz}
&&\epsilon_{abc}n^a \hat L^b z^c = \case12 \,{\rm sin}\,\bt_L \; 
{\rm sin}\,\bt_S\; 
{\rm sin}\,\bigl(\bph_L - \bph_S \bigr) \nonumber \\ 
&&- \case{\sqrt{3}}2 {\rm cos}\bph(t)\biggl(
 {\rm cos}\,\bt_L \; {\rm sin}\,\bt_S \; {\rm sin}\,\bph_S - 
 {\rm cos}\,\bt_S \; {\rm sin}\,\bt_L \;{\rm sin}\,\bph_L \biggr)
\nonumber\\
&&- \case{\sqrt{3}}2 {\rm sin}\bph(t)\biggl(
 {\rm cos}\,\bt_S \; {\rm sin}\,\bt_L \; {\rm cos}\,\bph_L - 
 {\rm cos}\,\bt_L \; {\rm sin}\,\bt_S \; {\rm cos}\,\bph_S \biggr)\;.\nonumber \\
&&
\end{eqnarray}

To recapitulate, Eqs.~(\ref{h_amp_phase})--(\ref{amp_phase}) 
give $h_I(t)$ in terms of $\bt_S$, 
$\bph_S$, $A_+$, $A_\times$, $F_+(t)$, and $F_\times(t)$.
$A_+$ and $A_\times$ are given by Eq.~(\ref{binary}), while 
the terms  $F_+(t)$, and $F_\times(t)$ are given in terms of
$\theta_S(t)$, $\phi_S(t)$, $\psi_S(t)$ by Eq.~(\ref{Fpc}), and finally
$\theta_S(t)$, $\phi_S(t)$, $\psi_S(t)$ are given by 
Eqs.~(\ref{cts}), (\ref{pts}), and (3.18)--(\ref{nlz}).
We note that an equivalent expression for $A_I(t)$ was 
derived independently by Giampieri~\cite{giampieri}.

\subsubsection{Two Detectors}
 
We have stated that with its three arms, LISA functions like
a pair of two-arm detectors, outputting
two linearly independent signals: 
$s_I(t) \equiv \biggl(\delta  L_1(t) - \delta L_2(t)\biggr)/L $
and  $s_{II'} \equiv \biggl(\delta L_2(t) - \delta L_3(f)\biggl)/L$. 
We now extend our above analysis to include both detectors.
We continue to assume the noise is stationary and Gaussian; nevertheless
the noise in different arms will generally be correlated. Let
$n_i(t) \equiv \delta L_i(t)/L$ be the noise in the $i^{th}$
arm; then
\begin{equation}
\langle \tilde n_i(f) | \tilde n_j(f') \rangle = C_{ij}(f)
\,\delta(f-f')
\label{coarm}
\end{equation}
\noindent where $C_{ij}(f) \ne 0$ in general.  Clearly then
the noise outputs of detectors $I$ and $II'$ will be correlated too.
However it is also clear that we can always find some
nontrivial $s_{II}$ which is a (frequency-dependent) linear 
combination $s_I$ and $s_{II'}$, and which is orthognal to $s_I$, in the
sense that the noise in $s_{II}$ is uncorrelated with the noise in $s_I$.
Just set 
\begin{equation}
\tilde s_{II} (f) = \tilde s_{II'} (f) - 
\tilde s_I(f){{ C_{12}(f)-C_{13}(f)-C_{22}(f)+C_{23}(f)}
\over{ C_{11}(f)-2C_{12}(f)+C_{22}(f)}}
\label{ortho}
\end{equation}
and then we have
\begin{equation}
\langle \tilde n_I (f) | \tilde n_{II} (f') \rangle = 0\;. 
\label{ortho}
\end{equation}
\noindent
We find thinking in terms of such orthogonal 
detectors to be very convenient for calculations.

Unfortunately, there do not yet exist estimates for how the
noise in LISA's three arms will
be correlated.  In this paper we will make the simplifying assumption that
the noise is `totally symmetric' among the three arms, 
in the following sense:
\begin{mathletters}
\label{symm}
\begin{eqnarray}
C_{12}(f) = C_{23}(f) = C_{31}(f) \;,
\label{s1}\\
C_{11}(f) = C_{22}(f) = C_{33}(f) \;.
\label{s2}
\end{eqnarray}
\end{mathletters}
It is not too unreasonable to suppose that the instrumental
noise will approximately totally symmetric,
since the individual satellites will all be virtually identical.
Also, one can easily show that the confusion noise due to an isotropic
background of gravitational wave sources must be totally symmetric.
The reason is that for an isotropic background 
$C_{ij}(f)/S_n(f)$ can only be a function of 
$(e_{ab}l_i^a l_j^b)^2$, where $e_{ab}$ is the Euclidean 3-metric.
The totally symmetric condition then follows from the facts that
LISA's arms are all the same length and the angle between 
any two different arms is $\pi/3$.  (Of course, our galaxy is not
spherically symmetric, and so the confusion noise from galactic 
binaries cannot be expected to satisfy the condition of total symmetry.)

For totally symmetric noise, the linear combination of 
$\tilde s_I(f)$ and $\tilde s_{II'}(f)$  
that is orthogonal to $\tilde s_I(f)$ is actually frequency-independent.
That combination is
\begin{eqnarray}
\label{sII} 
s_{II}(t) &=& 3^{-1/2}\biggl(\delta L_1(t) + \delta L_2(t) - 2\, \delta L_3(t)\,\biggr)/L \nonumber\\
&=& 3^{-1/2}\biggl(s_I(t) + 2 s_{II'}(t)\biggr)\;
\end{eqnarray}
\noindent 
which implies
\begin{equation}
\langle \tilde n_\alpha(f) | \tilde n_\beta(f') \rangle = \delta_{\alpha \beta}
\, \delta(f-f') \,S_n(f),
\label{nanb}
\end{equation}
\noindent where $\alpha$,$\beta$ take on values $I$ or $II$, and 
$S_n(f)$ is the spectral density for detector I.

In terms of the `unbarred' coordinates introduced in Sec.~III.A, it is
easy to show using Eq.~(3.1) that 
\begin{equation}
h_{II}(t)= \case{\sqrt{3}}2 \bigl(\case12 h_{xy} + \case12 h_{yx}\bigr)\; .
\label{h2t}
\end{equation}
\noindent That is, just as $h_I(t)$ is equivalent to the response of
of a $90^o$-interferometer (modulo the overall factor $\sqrt{3}/2$), 
so $h_{II}(t)$ is equivalent to the
response of {\it another} $90^o$-interferometer, rotated by $\pi/4$ radians
with respect to the first one. (This result was previously derived by
P. Bender, in unpublished work \cite{pbender_private}.)
It is therefore trivial to 
write down the beam coefficients for detector II:
\begin{mathletters}
\label{FII}
\begin{eqnarray}
F_{II}^+(\theta_S,\phi_S,\psi_S)  = F_I^+(\theta_S,\phi_S - \pi/4,\psi_S) \; 
\label{f2p}\\ 
F_{II}^{\times}(\theta_S,\phi_S, \psi_S)  =  F_I^{\times}(\theta_S,\phi_S - 
\pi/4, \psi_S)  \;
\end{eqnarray}
\end{mathletters}
Finally, in complete analogy with Eqs.~(\ref{h_amp_phase}) and (\ref{amp_phase}), we can write $h_{II}(t)$ as
\begin{equation}
h_{II}(t)= \case{\sqrt{3}}2 \, A_{II}(t) \cos [\int^t{2\pi f(t') dt'} + \varphi_{p,II}(t) 
+ \varphi_D(t)]\;.
\label{hampphase2}
\end{equation}
\noindent
where
\begin{mathletters}
\label{ampphase2}
\begin{eqnarray}
A_{II}(t) &=& \biggl(A_+^2 {F_{II}^+}^2(t)  + A_\times^2 {F_{II}^\times}^2(t) \biggr)^{1/2} \;,
\label{amp2}\\ 
\varphi_{p,II}(t) &=& \tan^{-1}\left({{-A_\times F_{II}^\times(t)}
\over {A_+ F_{II}^+(t)}}\right)\;,
\label{pphase2}
\end{eqnarray}
\end{mathletters}
\noindent and where $\varphi_D(t)$ is still given by Eq.~(\ref{Dphase}).

Figs.~3 and 4 show an example of the modulation patterns 
due to detector rotation, for one representative choice of
parameter values. 

\begin{figure}
\psfig{file=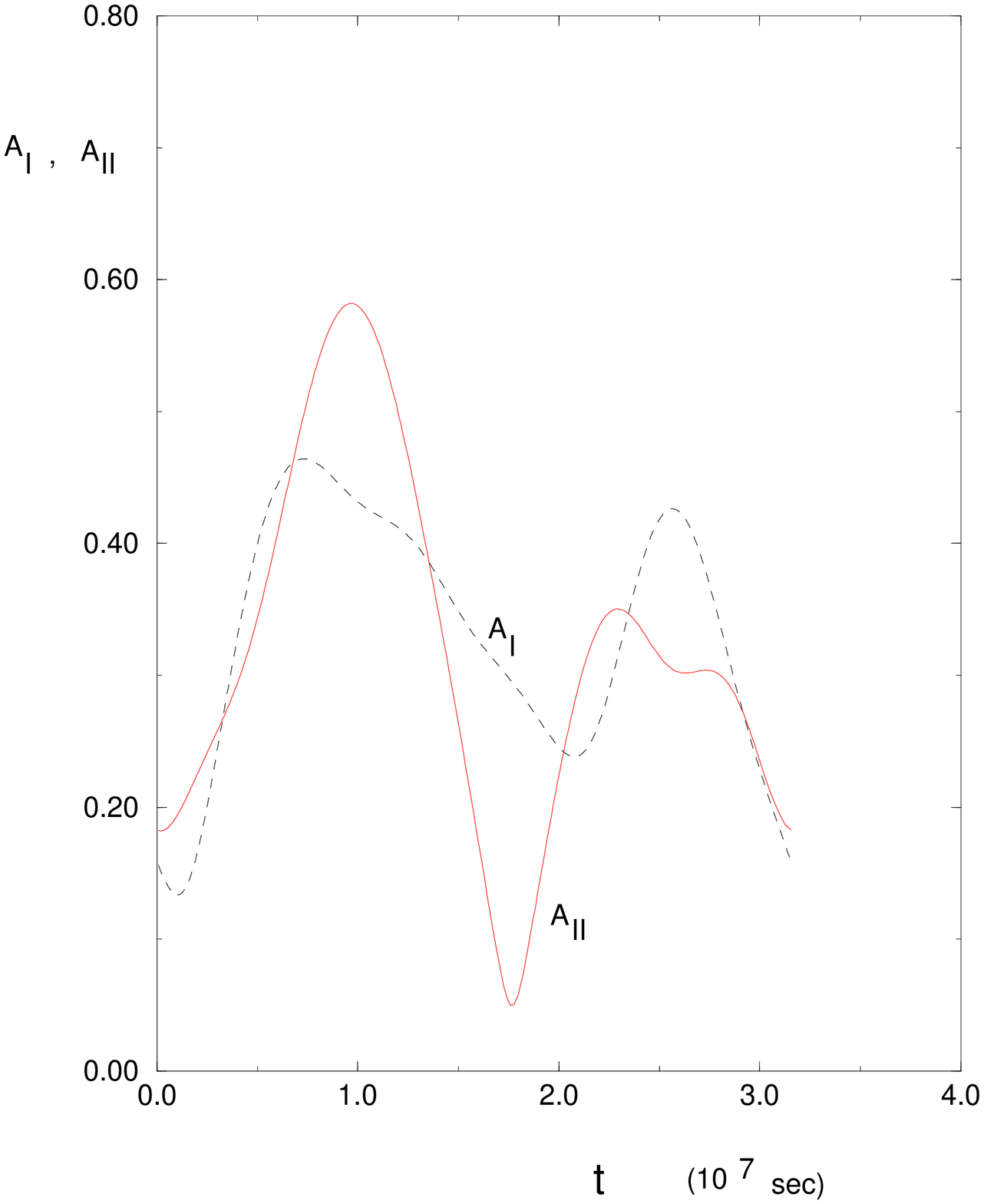,width=8cm,bbllx=18pt,bblly=144pt,bburx=592pt,bbury=718pt}
\vskip 1.5 cm
\caption{\label{figure3}  The amplitudes $A_I(t)$ and $A_{II}(t)$ during a
one-year observation, for the following choice of initial detector 
position and orientation and of the source's position and polarization:
$\bar \phi_0= 0, \alpha_0 = 0, \mu_S = 0.3, \phi_S = 5.0, \mu_L = -0.2, \phi_L = 4.0$.  The overall scale is arbitrary.}
\end{figure}

\begin{figure}
\psfig{file=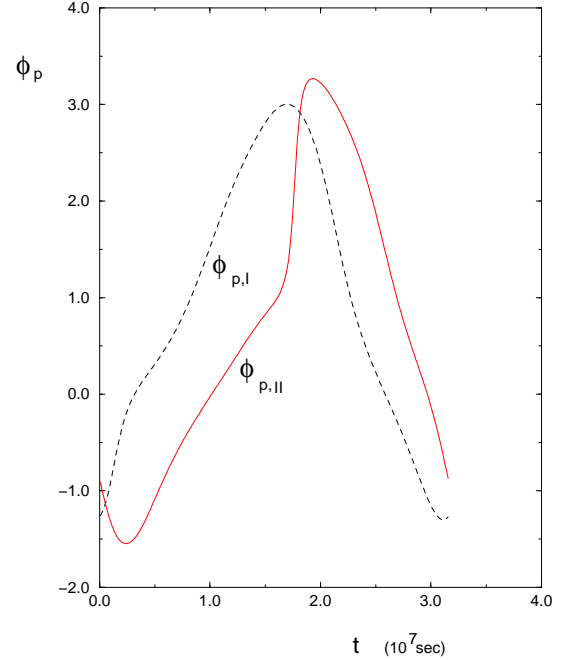,width=8cm,bbllx=18pt,bblly=144pt,bburx=592pt,bbury=718pt}
\vskip 1.5 cm
\caption{\label{figure4}  The polarization phases $\varphi_{p,I}(t)$ and 
$\varphi_{p,II}(t)$ during a one-year observation, for the same parameter values
as in Fig.~4.}
\end{figure}

\subsection{Noise Spectrum}

We will assume the noise spectra for detectors I and II are the 
same; both are given by $S_n(f)$, which we represent as the sum
of instrumental noise $S_{n,in}(f)$ and confusion noise $S_{n,co}(f)$:
\begin{equation}
\label{nsum}
S_n(f) = S_{n,in}(f) + S_{n,co}(f) \; .
\end{equation}
\noindent These two contributions are shown in Fig.~5.
We now consider them in turn.

\subsubsection{Instrument Noise}

It is a LISA design goal that the instrumental noise
be stationary and Gaussian; our analysis will assume that goal has been
met.
The following is the current best estimate of the spectral density of
the {\it instrumental} noise
$S_{n,in}(f)$ for detector I~\cite{Pre}.
\begin{equation}
S_{n,in}(f) = 5.049 \times 10^{5}\biggl[\alpha^2(f) +\beta^2(f) + 
\gamma^2\biggl]
\label{sn}
\end{equation}
\noindent where
\begin{eqnarray}
\label{abg}
\alpha(f) &=& 10^{-22.79}\, (f/10^{-3})^{-7/3} \;, \nonumber \\
\beta(f) &=& 10^{-24.54}\, (f/10^{-3}) \, ,
\ \ \ \gamma = 10^{-23.04}\;.
\end{eqnarray}
\noindent Here $\alpha(f)$ is mainly due to temperature fluctuations, 
$\beta(f)$ reflects the loss in sensitivity
when the gravitational wavelength becomes comparable to or shorter than
than the detector arm length, and $\gamma$ (a constant) 
is mainly due to photon shot noise. We will assume that Eqs.~(\ref{sn})--(\ref{abg}) above give the noise spectral density for detector II as well.

\subsubsection{Confusion Noise}

It seems very likely that gravitational waves from short-period, stellar-mass 
binaries will actually be more important than the instrumental noise in 
`drowning out' the signal from other types of sources. 
The issue of {\it why} stellar-mass binaries should be regarded
as effectively a noise source is a subtle but important one. 
We confine ourselves to a few remarks on this subject, and 
refer to the reader to Bender and Hils~\cite{bh} and Hils, Bender, and 
Webbink~\cite{Hils:1990} for further details.

The first remark is that the orbits of these stellar-mass binaries 
can be treated as Newtonian, and the radiation computed accurately from the
quadrupole approximation. The orbit of a Newtonian binary is 
periodic, so in Fourier space its gravitational radiation is composed
of discrete lines at $f = 2/P,\; 4/P,\; 6/P,\;$, etc., where $P$ is the orbital
period. (The sequence is multiples of $2/P$ instead of $1/P$ because the
radiation is quadrupolar.) It seems reasonable
to assume that most of the short-period, stellar-mass binaries 
have small eccentricity:
$e < 0.2$. In this case, $ > 60 \%$ of the power comes out in the lowest-frequency line, $f = 2/P$ ~\cite{Hils:1990}. 
Current estimates of the confusion noise therefore neglect
the higher-frequency lines.

Second, for an observation time of $T_0 \sim 1$ yr, the discrete
Fourier transform sorts monochromatic signals into frequency bins
of width  $\Delta f = 1/T_0 \sim 3 \times 10^{-8}$ Hz.  
The typical timescale on which these binaries evolve is $\agt 10^7$ yrs; 
thus in one year's observation, a binary's emitted GW frequency 
changes by $\alt f/10^7   = 10^{-10} (f/10^{-3}) $ Hz---i.e., much 
less than the width of one bin! 
Thus, ignoring for the moment the motion of the detector, each binary 
remains in the same bin throughout the observation.
The lower half of the LISA band, $10^{-4} - 10^{-3}$ Hz,
contains roughly $3 \times 10^4$ frequency bins, while
our galaxy contains $\sim 3 \times 10^7$ close white dwarf binaries
(CWDB's) with GW frequencies in this range, so roughly $10^3$ per bin.
Thus the problem of `fitting' for all the binaries, in order to then `take
them out' of the data, is extremely underdetermined: there are at least
$10^3$ times as many free parameters as there are data points.  
In fact, to model this signal accurately, one needs not only the frequency
of each source but its location and orientation, since 
the motion of the detector `smears out' the 
the signal over a frequency range $\sim 2/{\rm year}$, in a manner that
depends on these additional variables.  So the motion of the detector
only aggravates the problem of `fitting out' the stellar-mass binaries.
One might even argue that instrumental noise is in principle no different from
binary confusion noise; instrumental noise always arises from some
deterministic physical processes that one could also in principle 
model and then attempt to remove from the data by some fitting procedure
(or by monitoring these processes directly), but in practice one 
reaches a point where there are simply too many
variables--too many free parameters--to obtain a fit that has 
predictive power.

The number of galactic binaries per bin decreases with increasing 
frequency. Somewhere in the range $10^{-3}$ Hz to  $4 \times 10^{-3}$ Hz,
there is a transition from having many galactic binaries per bin to
having fewer than one, on average. At frequencies above this transition,
most of the information about some broad-spectrum source 
(such as a SMBH merger) will come from the bins that do not contain 
galactic binaries. At these frequencies, then, the binary confusion noise
is dominated by the extragalactic contribution. 

The following represents the current best estimate by 
Bender and Hils \cite{bh} for the level of the binary confusion noise:

\FL
\begin{equation}
\label{snc}
S_{n,co}(f) =\left\{ \begin{array}{ll} 
         10^{-42.685}\, f^{-1.9}    
& \mbox{ $f \le 10^{-3.15} $,} \nonumber \\
         10^{-60.325}\, f^{-7.5}   
          & \mbox{    $10^{-3.15} \le f \le 10^{-2.75} \,$}
             \nonumber \\
         10^{-46.85}\, f^{-2.6}
          & \mbox{ $ 10^{-2.75}\,\le f $ } \\
		\end{array}
	\right.
\end{equation}

\noindent where $f$ is measured in Hz. This estimate assumes a 
space density of CWDB's which is $10\%$ of the
theoretical value predicted by Webbink~\cite{webbink_1984}, 
and does not yet take into 
account the contribution from helium cataclysmic variables (which are
likely to be important in the range $1$ to $3$ mHz).

\begin{figure}
\psfig{file=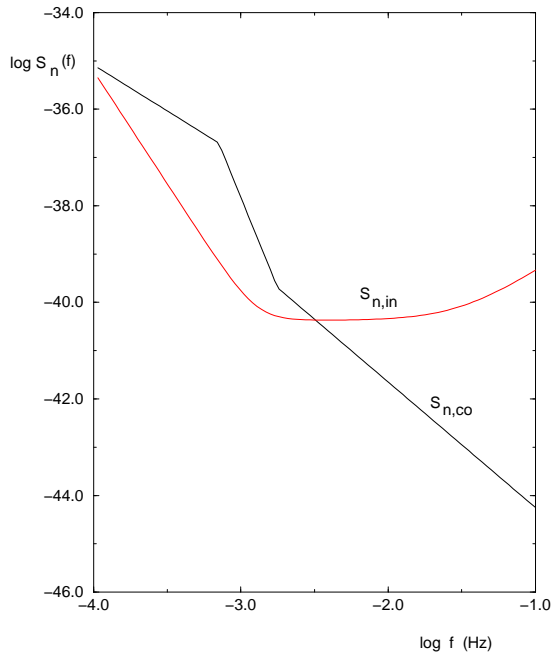,width=8cm,bbllx=18pt,bblly=144pt,bburx=592pt,bbury=718pt}
\vskip 1.5 cm
\caption{\label{figure4}  Shows the spectral density of both the
instrumental and confusion noise.  ${\rm log}_{10} S_{n,in}(f)$ and
${\rm log}_{10} S_{n,co}(f)$ are plotted versus $\rm{log}_{10}f$. The total noise
is given by $S_n(f) = S_{n,in}(f) + S_{n,co}(f)$.}
\end{figure}

\section{Monochromatic Sources}

Any monochromatic point source of gravitational waves {\it can be
thought of} as a circular-orbit binary: they are in 1-1 correspondence.
Therefore, in the rest of this section we shall always speak of
the angular resolution of LISA for circular-orbit binaries, but
the results would apply to any monochromatic source.  
An arbitrary circular-orbit binary is described by
{\it seven} free parameters: 
the frequency $f_0$ (as measured by an observer at the solar system
barycenter); the angles $\bt_S$, $\bph_S$, $\bt_L$, and $\bph_L$; 
an overall amplitude  proportional to ${\cal A} \equiv M_1 M_2/rD$; 
and a trivial overall phase $\varphi_0$ related to the choice of $t=0$. 

\subsection{The Measured Signal}

For a circular, Newtonian binary, the waveform $h_\alpha(t)$ 
($\alpha = I, II$) can be
written as
\begin{equation}
\label{monoh}
h_\alpha(t)= A_\alpha(t)\,{\rm cos}\,\chi_\alpha(t) 
\end{equation}
\noindent where 
\begin{equation}
\label{chi}
\chi_\alpha(t)= 2 \pi f_0 t + \varphi_0 + \varphi_{p,\alpha}(t) +  \varphi_D(t) \;.  
\end{equation}
\noindent where $\varphi_0$ is just a constant of integration, and
$A_\alpha(t)$, $\varphi_{p,\alpha}(t)$, and $\varphi_D(t)$ 
are given by Eqs.~(\ref{amp_phase}) and (\ref{ampphase2}).
  
The calculation of the Fischer matrix is simplified by the following trivial
observation. Although the measured frequency is not exactly constant, due
to the motion of the detector, it is very {\it nearly} the constant
$f_0$. Therefore
we can take the factor $1/S_n(f)$ out of the integral in Eq.~(\ref{inner}),
and write
\FL
\begin{mathletters}
\begin{eqnarray}
\label{ampphase}
\left(\partial_i {\bf h} \, |\, \partial_j {\bf h} \right)  &=& \case2{S_n(f_0)}
\, \sum_{\alpha = I,II} \;
\int_{-\infty}^{\infty} \, { \partial_i{\tilde h}_\alpha^*(f)
\partial_j{\tilde h}_\alpha(f)} \,\,  df
\label{int1}\\ 
&=& \case2{S_n(f_0)} \, \sum_{\alpha = I,II} \;
\int_{-\infty}^{\infty} \, \partial_i h_\alpha(t) \, \partial_j h_\alpha(t) dt \; .
\end{eqnarray}
\end{mathletters}
\noindent where we used Parseval's theorem to go from (\ref{int1}) 
to (4.3b).
Then using the fact that $f >>A^{-1} dA/dt$, we can approximate 
(4.3b) by 
\begin{eqnarray}
\label{final}
\left(\partial_i {\bf h} \, |\, \partial_j {\bf h} \right)  &=& 
S_n(f_0)^{-1}
\, \sum_{\alpha = I,II}\; \int_{-\infty}^{\infty}
\,\biggl[\partial_i A_\alpha(t) \partial_j A_\alpha(t) \nonumber \\
&+& A_\alpha^2(t)\partial_i \chi_\alpha(t)
\partial_j \chi_\alpha(t)\biggr] dt .\;
\end{eqnarray}
\noindent Thus to evaluate the Fisher 
matrix (\ref{final}), we need the derivatives of 
$A_\alpha(t)$ and $\chi_\alpha(t)$ with respect to the seven 
physical parameters
${\rm ln}\, {\cal A}$, $\varphi_0$, $f_0$, $\bt_S$, $\bph_S$, 
$\bt_L$, $\bph_L$.
Clearly one might straightforwardly use the
chain rule with Eqs.~(\ref{amp_phase}) and (\ref{ampphase2}) to determine the partial derivatives of
$A_\alpha(t)$ and $\chi_\alpha(t)$ 
with respect to the four angles 
$\bt_S$, $\bph_S$, $\bt_L$,and $\bph_L$, though the final expressions
would be cumbersome.  In our calculation, we preferred simply to take these  
derivatives numerically. The remaining partial derivatives
are:
\begin{mathletters}
\begin{eqnarray}
\label{dh} 
&&{ \partial A_\alpha(t) \over \partial {{\rm ln}\,{\cal A}}}  =   A_\alpha(t)\, ,\ \
{\partial A_\alpha(t) \over  \partial \varphi_0}  =   0\, ,\ \
{\partial A_\alpha(t) \over \partial f_0}  =   0\; ,\ \ \\
&&{ \partial \chi_\alpha(t) \over \partial {{\rm ln}\,{\cal A}}}  =   0\, ,\ \
{\partial \chi_\alpha(t) \over  \partial \varphi_0}  =   1\, ,\ \
{\partial \chi_\alpha(t) \over \partial f_0}  =   2 \pi t \;.
\label{dh1} 
\end{eqnarray}
\end{mathletters}

\subsection{Results}

For monochromatic sources, the detailed shape of the noise curve
has no bearing on the Fischer matrix; all that matters is $S_n(f_0)$,
where $f_0$ is the source frequency.  And $S_n(f_0)$ is inversely proportional
to the signal-to-noise of the detection, so we can eliminate $S_n(f_0)$
from the problem simply by normalizing the results to some fixed, fiducial
$S/N$.  In Table I, we normalize our results to $S/N = 10$, where $S/N$ is
the {\it total} signal-to-noise accumulated by both detectors I and II. 
The advantage of this way of representing our results is 
that Table I remains valid for {\it any} noise spectral density.  
(However the results do depend on our assumption that the noise is
totally symmetric among the three arms.)  Table I lists LISA's 
angular resolution
$\Delta \Omega_S$ for a one-year observation,  
for a range of source frequencies $f_0$ and for 
representative choices of angles $\bar\theta_S$, $\bar\phi_S$, $\bar\theta_L$, $\bar\phi_L$.
For these case, we also list $S_{I}/N$ and $\Delta \Omega_{S,I}$, the 
signal-to-noise and angular resolution of detector I taken alone.
(One expects $S_I/N$ to be roughly $2^{-1/2} S/N = 7.07$, but the exact 
value clearly depends on the various angles specified.) 
The sizes of the position error ellipses 
$\Delta \Omega_S$ and $\Delta \Omega_{S,I}$ 
simply scale like $(S/N)^{-2}$.

Since everything in the problem is periodic with period one year,
one obtains exactly the same results for $T$ years of observation,
when $T$ is an integer.  (That is, one obtains the same results after
normalizing to $S/N = 10$; if instead one normalizes to sources at
some fixed distance, then $S/N$ scales like $\sqrt{T}$ and
$\Delta \Omega_S$ scales like
$T^{-1}$.)   This scaling will hold approximately, but not exactly, 
when $T$ is some non-integer greater than $1$.  LISA's angular resolution 
would certainly be much worse for observation times significantly less 
than one year.

The results in Table I are easily summarized.  LISA's angular
resolution for monochromatic sources is roughly in the
range $10^{-3}$ to $10^{-1}$ steradians 
(equivalently, $3$ to $300$ square degress) for source frequencies in the
range $10^{-4} \le f_0 \le 10^{-2}$ and $S/N = 10$.
Having data from both detectors I and II provides hardly any improvement
in angular resolution, apart from the trivial improvement due to the
increased $S/N$.  Presumably this is because in one year's time, LISA's
changing orientation allows detector I by itself to measure 
both polarizations of the incoming wave fairly accurately.  LISA's
angular resolution is roughly a hundred times better at $10^{-2}$ Hz than
at $10^{-4}$ Hz (for fixed $S/N$).  Clearly this is because the Doppler
modulation is a much bigger effect in the higher-frequency sources. 

We note that one application of the results in this section is to compact 
stellar mass objects (like NS's) in orbit around SMBH's.  In that case,
orbital evolution is sufficiently slow on the timescale of 
the observation that the signal is effectively a sum of `lines' 
at frequencies $n/P$, where P is the orbital period.  To the extent that
the strongest line dominates, the source is therefore monochromatic.

\section{Supermassive Black Hole Mergers}

In this section we consider the information LISA could
extract from the collision of two supermassive black holes.
Note that since high signal-to-noise ratios are expected for
this case, we expect the Fischer matrix approach to error estimation
to be quite accurate.

SMBH collisons are related by simple re-scaling to the 
mergers of stellar-mass black holes; many of the issues that arise
in GW data analysis are identical for the two cases, so we take
take advantage of the extensive literature on the latter.
Our exposition in this section will therefore
be highly abbreviated; we refer to \cite{cutler_flanagan} for more 
extensive discussion and derivations.  

Unlike our treatment of monochromatic sources in Sec.~IV, our
treatment here will be based on two simplifying assumptions that
probably are not justified physically.  Most importantly, we
neglect the spin-induced precession of the binary's orbital
plane, and its corresponding modulation of the waveform 
amplitude and phase. Given that the rotation of the detector and
the rotation of the binary's orbital plane modulate the signal in
very similar ways, it may in practice prove difficult to disentangle
these two effects in the data analysis, and the accuracy of
position measurement would be correspondingly degraded. 
The second complication we omit is the possible eccentricity of
the orbit. 
We plan to explore these two complications in later work.

One might expect such collisions to occur at significant redshift
($z \ge 1$) so in this section we carry along the factors of $(1+z)$ that we
have ignored in earlier parts of the paper (see Ref.~\cite{draza}).  
These collisions would be
visible to LISA for a significant fraction of year.  Assuming  
circular orbits, the time interval from the instant $t(f)$ that
quadrupole frequency sweeps past $f$ until the instant $t_c$ when
the two bodies collide and merge is (to lowest order):

\begin{eqnarray}
\label{tf}
t_c - t(f) & = & 5(8\pi f)^{-8/3}\biggl({\cal M}\,(1+z)\biggr)^{-5/3}
\nonumber \\
& = & 3.003 \times 10^6 {\rm s}\, \bigl(f/10^{-4}\bigr)^{-8/3}\biggl(
\case{{\cal M}\,(1+z)}{10^6 M_\odot}\biggr)^{-5/3}
\end{eqnarray}
\noindent where ${\cal M} = M_1^{3/5} M_2^{3/5}/(M_1 + M_2)^{1/5}$,
and $f$ and $t$ are the frequency and time measured by an observer at the
Earth.

\subsection{The Measured Signal}

We define a 
signal $H(t)$ by
\begin{equation}
H(t) \equiv \case{2\, M_1\, M_2 (1+z)}{D_L\, r(t)} \;
\cos \;\int_0^t{f(t') \; dt'}
\label{carrier}
\end{equation}
where $f(t)$ is the (redshifted) frequency a detector {\it would} measure 
if it were nonrotating and
its position were fixed at the Solar system barycenter.
In Eq.~(\ref{carrier}), $M_1$, $M_2$ and $r$ are the unredshifted masses and orbital separation
that would be measured by an observer near the source, and $D_L$
is the `luminosity distance' to the source~\cite{draza}.
We think of $H(t)$ as a `carrier' signal, which is modulated by
the motion of the detector.  The {\it measured} signal $h_\alpha(t)$
is given by
\begin{equation}
h_\alpha(t) = \Lambda_\alpha(t)\; \cos \;\biggl( \int_0^t{f(t') \; dt'
+ \varphi_{p,\alpha}(t) + \varphi_D(t) \biggr)} 
\label{modsmbh}
\end{equation}
\noindent where $\Lambda_\alpha(t)$, defined by
\begin{equation}
\Lambda_\alpha(t) 
\equiv \case{D_L\, r(t)}{2\, M_1\, M_2 (1+z)} A_\alpha(t)\;,
\label{lamdef}
\end{equation}
\noindent basically encodes the amplitude modulation.
How is $\tilde h_\alpha(f)$ related to $\tilde H(f)$, 
the Fourier transform
of the carrier? Since $\Lambda_\alpha(t)$, 
$\varphi_{p,\alpha}(t)$, and $\varphi_D(t)$ 
all vary on timescales of $\sim 1 {\rm yr} >> 1/f$, we can
approximate $\tilde h_\alpha(f)$ using the 
stationary phase approximation. This gives \cite{haris}
\begin{equation}
\tilde h_\alpha(f) = \Lambda_\alpha(t) e^{-i\bigl( \varphi_{p,\alpha}(t) +  \varphi_D(t)\bigr)}  \tilde h(f) \; 
\ \ \ (f>0)
\label{stat}
\end{equation}
where $t = t(f)$ is the instant at which the GW frequency
sweeps through the value $f$. So given  $\tilde H(f)$, we have
a simple (approximate) expression for $\tilde h_\alpha(f)$.

In this paper, we use a `model' of $\tilde H(f)$ that was
developed in \cite{3min,cutler_flanagan}. This model is
based on the so-called `restricted PN approximation'.  
In brief, the model includes post-Newtonian 
corrections to the {\it phase} of the waveform through ${\cal O}(v/c)^3$,
but it takes the  waveform to 
be quadrupolar, with an overall amplitude given by the lowest-order
approximation. A more detailed explanation of this model is given
in \cite{cutler_flanagan}. Specifically, our model $\tilde H(f)$ is
\FL
\begin{equation}
\label{pnshf}
\tilde H(f) = \left\{ \begin{array}{ll}
{\cal A}\, f^{-7/6} e^{i \Psi}
 &\ \ \  \mbox{$0 < f < f_{cut-off}$}\nonumber \\
0
&\ \ \  \mbox{$f>f_{cut-off}$}\\
		\end{array} 
\right. 
\end{equation}
where 
\FL
\begin{eqnarray}
\label{pnspsi}
\cal A & = & (5/96)^{1/2} \pi^{-2/3} D_L^{-1} \bigl[{\cal M} (1+z)\bigr]^{5/6}
\;  \label(cA) \\
\Psi(f) & = &
2\pi f t_c -\phi_c -\pi/4 +{3\over 4}\bigl(8 \pi {\cal M} (1+z) f
\bigr)^{-5/3} \nonumber \\
\mbox{} & \times & \, \biggl[1+ {20\over 9}\left({743\over 336}+{{11
\mu}\over {4M}}\right)x +\left(4\beta -16 \pi \right)x^{3/2} \biggr] \; . 
\end{eqnarray}
\noindent Here $M \equiv M_1 + M_2$, $\mu \equiv M_1 M_2/M$, ${\cal M} 
\equiv \mu^{3/5} M^{2/5}$, and the PN expansion parameter $x(f)$ 
is defined by 
\begin{equation}
\label{xf}
x(f) \equiv \biggl(\pi M (1+z) f \biggr)^{2/3} \; .  
\end{equation}
\noindent The parameters $t_c$ and $\phi_c$ are just 
constants of integration. The term $\beta$ is a $P^{1.5}N$ 
spin-orbit coupling term defined in \cite{cutler_flanagan}. 
($\beta$ is only approximately conserved by the
$P^{1.5}N$ equations of motion, but in our model we treat it as a 
constant). In this paper, we `cut-off' the signal (somewhat arbitarily) at
$f_{cut-off} = (3^{3/2} \pi M (1+z) )^{-1}$, corresponding to $r/M=3$.
In fact, the value of this cut-off substantially affects the
calculated signal-to-noise $S/N$ in cases where 
$M (1+z)$ is greater than $\sim 3 \times 10^6 M_\odot$,
because in these cases  most of the $S/N$ accumulates at the very 
end of inspiral.  That is because LISA's noise curve $S_n(f)$ falls 
very rapidly with increasing $f$ for $f <  3 \times 10^{-3}$ Hz.   
However we will find that 
the predicted angular resolution of the detection, for a source at
fixed distance, is rather insensitive to $f_{cut-off}$.
 
From Eq.~(\ref{stat}) we therefore have
\begin{equation}
\tilde h_I(f) = \Lambda\bigl(t\bigr) e^{i\bigl(\chi(f)- \varphi_p(t) -  
\varphi_D(t)\bigr)}  \tilde h(f) \; 
\ \ \ (f>0)
\label{h1}
\end{equation}
\noindent where $t=t(f)$ is given through ${\cal O}(v/c)^3$ by 
\cite{cutler_flanagan}
\FL
\begin{eqnarray}
\label{pntf}
t(f) & = & t_c - 5(8\pi f)^{-8/3} \bigl[{\cal M}(1+z)\bigr]^{-5/3}
\biggl[1 + \nonumber \\
\mbox{} & & {4\over 3}\left({{743}\over{336}} + 
{{11\mu}\over{4M}}\right)x - {{32\pi}\over 5} x^{3/2} +  O(x^2) \biggr] \;. 
\end{eqnarray}
 
Note that our model of the signal is the just inspiral waveform:
it does not include the final merger and ringdown.  Partly 
this is necessitated by our current ignorance about the final merger.
However it also seems to us that this neglect is justified by
the particular problem we are trying to solve; that is, it seems unlikely
that signal from the final burst will significantly improve
LISA's angular resolution, even if it dominates the signal-to-noise. 
The reason is
that the final burst is only a few seconds long, and obviously 
LISA's orientation and velocity hardly change in that interval, so there's
modulational encoding of the source position.
(Of course,
the final burst could give extra information about the chirp mass $\cal M$,
which is correlated with the position unknowns, but 
$\cal M$ is already very well determined by the lower-frequency data.)

Note that $\tilde h_\alpha(f)$ depends on 9 physical parameters:
${\cal M}$, $\mu$, $\beta$,
$\phi_c$, $t_c$, ${\rm ln}\,D_L$, $\bt_S$, $\bph_S$, $\bt_L$, $\bph_L$.
The next step is to evaluate the Fisher matrix~(\ref{bardx}).
As in Sec.~IV, we numerically evaluate the
partial derivatives of $\tilde h_\alpha(f)$ with respect to the four angles
$\bt_S$, $\bph_S$, $\bt_L$, $\bph_L$, using Eqs.~(\ref{amp_phase}) and (\ref{ampphase2}).
The partial derivatives of $\tilde h_\alpha(f)$ with respect to the
other 5 parameters are given by Eqs.~(3.14) and (3.19) 
of \cite{cutler_flanagan}.

\subsection{Results}

Using Eq.~(4.3) as our model for the signal, we have computed 
the variance-covariance
matrix $\Sigma^{ij}$ for a wide range of masses and angles.  
We note that since the signal-to-noise is high for SMBH mergers, 
the Fischer matrix approximation, Eq.~(\ref{bardx}), is expected
be quite accurate.

Clearly there is a very large, non-trivial parameter space to explore:
${\cal M}$, $\mu$, $\beta$, $\bt_S$, $\bph_S$, $\bt_L$, $\bph_L$.
(The Fisher matrix is independent of $\phi_c$ and $t_c$, and ${\rm ln}\, D_L$
just affects the overall scaling.)
Here we will look at just a few representative cases. Our results
are shown in Table II.  In all cases we take $\beta = 0$,
(that is, the {\it true} value of $\beta$ is zero, but the
best-fit value can be non-zero), and we take the initial position and
orientation of the detector to be $\bar\phi_0= 0, \alpha_0= 0$.
We take as our fiducial source a binary at $z=1$ in a low-density
($\Omega_0 = 0$) universe with $H_0 = 75$ km/s-Mpc; consequently our fiducial 
distance is $D_L = H_0^{-1} = 12.253 \times 10^{9}$ yr. 
The masses listed in Table II are the `true' masses, as they appear in
Eq.~(5.2), not their redshifted versions.  All results are for one year
of data.

As stated earlier, for ${\cal M} > 3 \times 10^{6} M_\odot$,
{\it most} of the signal-to-noise
accumulates at the very end of the inspiral, and therefore the
total $S/N$ we calculate with our model waveform 
depends rather sensitively on how one assumes the signal 
is `cut-off' as the two bodies merge.  Therefore the $S/N$ results 
listed in Table II should not be regarded as accurate to better 
than a factor of $\sim 2$.  However we find that our results 
for $\Delta \Omega_S$ do {\it not} depend sensitively on this cut-off
(for a given source at fixed distance).  We also find that, for
the mass range we looked at ($10^4$ to $10^7 M_\odot$), increasing
the observation to include longer than the final year did not
significantly increase the angular resolution.

The following points emerge from Table II.  
Unlike the case for monochromatic sources, having essentially two 
detectors {\it does} substantially increase LISA's angular resolution for
SMBH mergers. The angular resolution 
$\Delta \Omega_S$ 
achievable by detectors I and II combined is
roughly $10^{-4}$ steradians. The angular resolution depends strongly on the masses and 
the particular angles involved, however.  $\Delta \Omega_S$ is
roughly in the range $10^{-5}-10^{-3}$ steradians for 
$10^5 M_\odot < {\cal M}(1+z) < 10^7 M_\odot$. $\Delta \Omega_S$ is
generally larger for the lower mass black holes (${\cal M} \approx 10^4-10^5
M_\odot$)  because the signal-to-noise is generally smaller in these cases.

The angular resolution achievable by detectors I and II combined 
is roughly an order of magnitude better than that 
achievable with detector I alone.  Notice this is quite different from 
the case of monochromatic sources, where the improvement was only
a factor of $\sim 2$. It seems clear that this
difference arises because, in the SMBH case,
the time-scale over which most of the signal-to-noise is accumulated is 
rather shorter than a year. During this `effective' integration time
the orientation of detector I does not change dramatically, 
and so detector I is effectively
sensitive to only a single polarization.

The distance determination
accuracy $\Delta D_L/D_L$ for SMBH mergers is roughly 
in the range $0.1\%-30\%$, with $\sim 1\%$ being typical.
This is much
worse than the naive guess of $\Delta D_L/D_L \approx (S/N)^{-1}$.
Clearly the `extra error' is due to correlations between $D_L$ and the various angles describing the source.
The quantity $\Delta \Omega_L  \equiv 2\pi\,\biggl[\Delta {\bar\mu}_L \, \Delta {\bar\phi}_L - \left< \Delta {\bar\mu}_L \, \Delta {\bar\phi}_L \right> \,\biggr]$ represents LISA's  accuracy in determining the orbital plane of the binary.
We find that large values
of $\Delta D_L/D_L$ have a strong positive correlation with large values of
$\Delta \Omega_L$, as one might expect.

Finally we note that a recent paper by Schilling~\cite{schilling} has
looked more carefully LISA'a transfer function, and concludes that
the high-frequency part of the instrumental noise is somewhat lower than
previously estimated, so that the term $\beta(f)$ in Eqs.~(\ref{sn})--(\ref{abg}) should
be reduced by some factor $\alt 1.5$.  Re-running the code with this change,
we find, makes a negligible correction to the values of $\Delta\Omega_S$ listed
in Table II.

\section{Conclusions and Future Work}

  We have seen that, for SMBH mergers, LISA should achieve an angular 
resolution of (very roughly) $\sim 0.3 $ square degrees.  
What are the implications of this result for the idea of using such mergers to determine the cosmological parameters $H_0$, 
$\Omega_0$, and $\Lambda_0$?  Since LISA can determine the luminosity
distance to the source but not its redshift, one clearly needs the reshift
of the host galaxy or galactic cluster to do cosmology.  Since one square
degree contains $\sim 10^4$ $L_*$ galaxies, the LISA 
measurement alone is clearly insufficient to identify the host galaxy
or cluster.  On the other hand: because the signal-to-noise is so large,
one will know that a merger is occurring weeks before the final burst.
We have checked that, more than a day before the final burst, LISA will have
achieved most of the angular resolution indicated in Table II.
Thus one will know very accurately {\it when} the final burst will occur,
and will know to within a degree {\it where} it will occur.
At the right time, every available telescope can be trained at the right
spot in the sky (as happened with the impact of the comet 
Shoemaker-Levy on Jupiter), and, if one is lucky, the merging binary 
could `send up a flare' electromagnetically~\cite{bbr}. Of course, 
a flare is possible only if there is normal matter involved in the 
collison; e.g., if at least one of the black holes has preserved an
accretion disk up to the point of merger.  It is possible there
could be some electromagnetic signal even before the final burst.
Clearly, these possibilities deserves more investigation.

  Finally, we list some ways in which our analysis could be improved.
First, it would be useful to have a better understanding of how the 
instrumental  noise in the different arms will be correlated;
our assumption of `total symmetry' between the different arms was a crude estimate intended just to get us started. Of course it would also be useful to have a better estimate of the confusion noise levels, but it may take a working LISAto provide that!  Second, our treatment of monochromatic sources could
be improved by doing a full Monte Carlo estimation of the errors.  
The S/N for any monochromatic source is likely to be low, it is not clear
how reliable our Fischer matrix approxiamtion is in this case.
Thirdly, one could clearly do a much more systematic analysis of
the parameter space: the parameters in Table II were chosen more
or less at random. 
Finally, a better treatment of the SMBH case would allow for eccentric
orbits and spin-induced precession of the binary's orbital plane.  
Work to incorporate these last two effects in our analysis is now in 
progress~\cite{alberto_inprogress,alberto_curt}.

\acknowledgments I thank Giacomo Giampieri, Lee Lindblom, 
Bernard Schutz, Tuck Stebbins, and Alberto Vecchio for helpful discussions.
Special thanks are due to Peter Bender and Dieter Hils for 
several discussions of LISA's noise sources and for showing me their
recent work on confusion noise 
in advance of publication~\cite{bh}.
This work was supported by NSF grant PHY-9507686 and by an 
Alfred P. Sloan Fellowship, and was largely carried out while I was a visitor
at the Max Planck Institute for Gravitational Physics in Potsdam.

\vskip 2cm 
\begin{table}
\caption{LISA's angular resolution $\Delta\Omega_S$ (in steradians)
for monochromatic sources.
Results are for a 1-year observation, with source strength normalized so that total (i.e., detectors I and II combined) $S/N = 10$.  
$S_I/N$ and $\Delta \Omega_{S,I}$ are the signal-to-noise and
angular resolution achievable by detector I alone, for 
the same source strength. LISA's initial position and orientiation are given by
$\alpha_0 = \bar\phi_0 =0$.
\label{table1}}
\begin{tabular}{llllllll}
$f_{GW}$&$\bar\mu_S$&$\bar\phi_S$&$\bar\mu_L$&$\bar\phi_L$&$S_I/N$&$\ \ \Delta\Omega_{S,I}$&$\ \ \Delta\Omega_S$\\
\tableline
$10^{-4}$& \ 0.3 & 5.0 & -0.2 & 4.0  & 7.07 &$1.89 \times 10^{-1}$ & $7.79 \times 10^{-2}$\\ 
$10^{-4}$& \ 0.3 & 5.0 & \ 0.2 &  0.0  & 7.19 &$1.87 \times 10^{-1}$ &$ 7.41 \times 10^{-2}$\\ 
$10^{-4}$& -0.3 & 1.0 & -0.2 & 4.0  & 6.89 &$1.17 \times 10^{-1}$ &$ 7.10 \times 10^{-2}$\\ 
$10^{-4}$& -0.3 & 1.0 & \ 0.8 &  0.0  & 6.80 &$1.26 \times 10^{-1}$ & $7.10 \times 10^{-2}$\\
\tableline
$3\times 10^{-4}$&\  0.3 & 5.0 & -0.2 & 4.0  & 7.07 &$1.47 \times 10^{-1}$ & $6.41 \times 10^{-2}$\\
$3\times 10^{-4}$&\  0.3 & 5.0 &\  0.2 &  0.0  & 7.19 &$1.41 \times 10^{-1}$ & $6.15 \times 10^{-2}$\\ 
$3\times 10^{-4}$& -0.3 & 1.0 & -0.2 & 4.0  & 6.89 &$1.04 \times 10^{-1}$ &$ 6.20 \times 10^{-2}$\\ 
$3\times 10^{-4}$& -0.3 & 1.0 &\ 0.8 &  0.0  & 6.80 &$1.17 \times 10^{-1}$ &$ 
6.28 \times 10^{-2}$\\
\tableline
$10^{-3}$&\  0.3 & 5.0 & -0.2 & 4.0  & 7.07 &$6.15 \times 10^{-2} $&$ 2.91 \times 10^{-2}$\\
$10^{-3}$&\  0.3 & 5.0 &\  0.2 &  0.0  & 7.19 &$6.04 \times 10^{-2}$ & $2.87 \times 10^{-2}$\\ 
$10^{-3}$& -0.3 & 1.0 & -0.2 & 4.0  & 6.89 &$6.02 \times 10^{-2}$ &$ 3.17 \times 10^{-2}$\\ 
$10^{-3}$& -0.3 & 1.0 &\  0.8 &  0.0  & 6.80 &$6.85 \times 10^{-2}$ & $3.10 \times 10^{-2}$\\
\tableline
$3\times 10^{-3}$&\  0.3 & 5.0 & -0.2 & 4.0  & 7.07 &$1.50 \times 10^{-2}$ & 
$ 7.23 \times 10^{-3}$\\
$3\times 10^{-3}$&\  0.3 & 5.0 &\  0.2 &  0.0  & 7.19 &$1.58 \times 10^{-2}$ &
$ 7.41 \times 10^{-3}$\\ 
$3\times 10^{-3}$& -0.3 & 1.0 & -0.2 & 4.0  & 6.89 &$1.71 \times 10^{-2}$ &
$ 7.60 \times 10^{-3}$\\ 
$3\times 10^{-3}$& -0.3 & 1.0 &\  0.8 & 0.0  & 6.80 &$1.75 \times 10^{-2}$ &
$ 7.04 \times 10^{-3}$\\
\tableline
$10^{-2}$&\  0.3 & 5.0 & -0.2 & 4.0  & 7.07 &$1.93 \times 10^{-3}$ & 
$9.05 \times 10^{-4}$\\ 
$10^{-2}$&\  0.3 & 5.0 &\  0.2 &  0.0  & 7.19 &$2.16 \times 10^{-3}$ &
$ 9.55 \times 10^{-4}$\\ 
$10^{-2}$& -0.3 & 1.0 & -0.2 & 4.0  & 6.89 &$1.97 \times 10^{-3}$ & 
$7.98 \times 10^{-4}$\\ 
$10^{-2}$& -0.3 & 1.0 &\  0.8 &  0.0  & 6.80 &$1.95 \times 10^{-3}$ & 
$7.60 \times 10^{-4}$\\ 
\end{tabular}
\end{table}

\newpage
\widetext
\begin{table}
\caption{LISA's angular resolution for SMBH mergers.
All mergers are taken to occur at redshift $z=1$ and 
luminosity distance $D_L = H_0^{-1}$, where
we take $H_0 = 75$ km/s-Mpc.  Results marked with subscript ``I''
are for detector I alone; results without a subscript represent
the signal-to-noise and acuracies achievable using both detectors I and II.
LISA's initial position and orientiation are given by
$\alpha_0 = \bar\phi_0 =0$.
\label{table2}}
\begin{tabular}{ccddddddddddd}
$M_1$ & $M_2$&$\bar\mu_S$ &$\bar\phi_S$&$\bar\mu_L$&$\bar\phi_L$&$S_{I}/N\ \ $&$ S/N\  $&
$\, \Delta\Omega_{S,I}$&$\,\ \ \Delta\Omega_S$
&${{\Delta D_L}/{D_L}}$&\,${{\Delta \mu}/{\mu}}$  \\
$(M_\odot) $ & $(M_\odot)$&$$ &$$&$$ &$$&$$&$$&($10^{-5}$ str)&($10^{-5}$ str) 
&$ (\times 10^{-2})$&$(\times 10^{-2})$  \\
\tableline
$10^7$ & $10^7$ &  0.3 & 5.0 &0.8 & 2.0  & 975. & 1337. & 151.  &1.50 & 1.51   &1.51   \\
$10^7$ & $10^7$ &  -0.1 & 2.0 & -0.2 & 4.0  & 1436. & 2086. &252.  &15.3
&1.34   &1.09   \\
$10^7$ & $10^7$ &  -0.8 & 1.0 & 0.5 & 3.0  & 3152. & 4909. & 146.  &28.7 
&0.299   &0.627   \\
$10^7$ & $10^7$ &  -0.5 & 3.0 & -0.6 & -2.0  & 2506. & 3363. &230.  &
24.1& 0.935   & 0.842  \\ 
$10^7$ & $10^7$ &  0.9 & 2.0 & -0.8 & 5.0  & 4612. & 6179. & 53.0  &12.8
& 15.7     & 0.452    \\
$10^7$ & $10^7$ &  -0.6 & 1.0 &0.2 & 3.0  & 2387.  & 3943. & 161.  &
38.2& 0.535    & 0.772    \\
$10^7$ & $10^7$ &  -0.1 & 3.0 & -0.9 & 6.0  & 3413. & 3986. &451.  &
77.3& 0.825    & 0.803    \\   
\tableline
$10^7$ & $10^6$ & 0.3 & 5.0 &0.8 & 2.0  & 469. & 641. & 124.  &1.51 
& 1.31    & 0.513    \\
$10^7$ & $10^6$ &  -0.1 & 2.0 & -0.2 & 4.0  & 687. & 1001. &185.  &12.2
&1.15   &0.375   \\
$10^7$ & $10^6$ &  -0.8 & 1.0 &0.5 & 3.0  & 1483. & 2310. &102.  &20.6 
&0.277   &0.219   \\
$10^7$ & $10^6$ &  -0.5 & 3.0 & -0.6 & -2.0  & 1181. & 1588. &152.  &17.8&0.825   &0.298   \\  
$10^7$ & $10^6$ & 0.9 & 2.0 & -0.8 & 5.0  & 2192. & 3187. &42.2 &9.49
&13.7    &0.159   \\
$10^7$ & $10^6$ &  -0.6 & 1.0 &0.2 & 3.0  & 1125.  & 1852. &115. &27.0&0.468   &0.269   \\
$10^7$ & $10^6$ &  -0.1 & 3.0 & -0.9 & 6.0  & 1611. & 1883. &308. &55.4&0.691   &0.284   \\  
\tableline
$10^6$ & $10^6$ & 0.3 & 5.0& 0.8 & 2.0  & 2771. & 3803. & 93.0  &0.86 
&1.16   &0.319   \\
$10^6$ & $10^6$ &  -0.1 & 2.0 & -0.2 & 4.0  & 4088. & 5930. &126.  &9.05
&1.01   &0.228   \\
$10^6$ & $10^6$ &  -0.8 & 1.0 &0.5 & 3.0  & 9009. & 14030. &70.4  &15.4 &0.212   &0.127   \\
$10^6$ & $10^6$ &  -0.5 & 3.0 & -0.6 & -2.0  & 7160. & 9599. &103.  &13.7 &0.711   &0.172   \\ 
$10^6$ & $10^6$ & 0.9 & 2.0 & -0.8 & 5.0  & 13142. & 19157. &24.4 &6.85
&11.8    &0.091   \\
$10^6$ & $10^6$ &  -0.6 & 1.0 &0.2 & 3.0  & 6820.  & 11271. &76.0 &20.5 &0.392   &0.156   \\
$10^6$ & $10^6$ &  -0.1 & 3.0 & -0.9 & 6.0  & 9741. & 11376. &217. &39.4 &0.562   &0.159   \\   
\tableline 
$10^6$ & $10^5$ & 0.3 & 5.0 &0.8 & 2.0  & 1317. & 1807. &151.  &1.67 
&1.55   &0.107   \\
$10^6$ & $10^5$ &  -0.1 & 2.0 & -0.2 & 4.0  & 1942. & 2818. &163.  &16.5
&1.35   &0.077   \\
$10^6$ & $10^5$ &  -0.8 & 1.0 &0.5 & 3.0  & 4277. & 6661. &99.3  &25.1 &0.274   &0.045   \\
$10^6$ & $10^5$ &  -0.5 & 3.0 & -0.6 & -2.0  & 3400. & 4558. &134.  &23.7&0.940   &0.061   \\
$10^6$ & $10^5$ & 0.9 & 2.0 & -0.8 & 5.0  & 6242. & 9098. &41.6 &10.5
&14.3    &0.031   \\
$10^6$ & $10^5$ &  -0.6 & 1.0 &0.2 & 3.0  & 3238.  & 5352. &109. &34.4 &0.506   &0.055   \\
$10^6$ & $10^5$ &  -0.1 & 3.0 & -0.9 & 6.0  & 4625. & 5402. &288. &59.7 &0.649   &0.055   \\   
\tableline
$10^5$ & $10^5$ & 0.3 & 5.0 &0.8 & 2.0  & 667. & 913. &294.  &4.54 
&2.49   &0.199   \\
$10^5$ & $10^5$ &  -0.1 & 2.0 & -0.2 & 4.0  & 981. & 1425. &332. &42.5
&2.18   &0.145   \\
$10^5$ & $10^5$ &  -0.8 & 1.0 &0.5 & 3.0  & 2156. & 3358. &239. &62.5 &0.436   &0.086   \\
$10^5$ & $10^5$ &  -0.5 & 3.0 & -0.6 & -2.0  & 1714. & 2300. &313.  &61.6 &1.51   &0.118  \\  
$10^5$ & $10^5$ & 0.9 & 2.0 & -0.8 & 5.0  & 3152. & 4593. &89.9 &24.1
&20.9    &0.059   \\
$10^5$ & $10^5$ &  -0.6 & 1.0 &0.2 & 3.0  & 1633.  & 2697. &269. &91.1&0.813   &0.105   \\
$10^5$ & $10^5$ &  -0.1 & 3.0 & -0.9 & 6.0  & 2334. & 2725. &648. &144.&0.978   &0.104   \\  
\tableline
$10^5$ & $10^4$ & 0.3 & 5.0 &0.8 & 2.0  & 238. & 323. &643.  &14.3 
&4.28   &0.120   \\
$10^5$ & $10^4$ &  -0.1 & 2.0 & -0.2 & 4.0  & 348. & 508. &944. &118.
&3.78   &0.084   \\
$10^5$ & $10^4$ &  -0.8 & 1.0 &0.5 & 3.0  & 751. & 1171. &849. &198. &0.788   &0.049   \\
$10^5$ & $10^4$ &  -0.5 & 3.0 & -0.6 & -2.0  & 599. & 807. &1088.  &187.&2.58   &0.068   \\
$10^5$ & $10^4$ & 0.9 & 2.0 & -0.8 & 5.0  & 1115. & 1619. &281. &69.1
&33.3    &0.035   \\
$10^5$ & $10^4$ &  -0.6 & 1.0 &0.2 & 3.0  & 570.  & 939. &1199. &317.&1.48   &0.057   \\
$10^5$ & $10^4$ &  -0.1 & 3.0 & -0.9 & 6.0  & 820. & 958. &2257. &489.&1.94   &0.061   \\  
\tableline
$10^4$ & $10^4$ & 0.3 & 5.0 &0.8 & 2.0  & 109. & 148. &684.  &19.2 
&3.11   &0.343   \\
$10^4$ & $10^4$ &  -0.1 & 2.0 & -0.2 & 4.0  & 156. & 233. &1293. &85.5
&2.69   &0.250   \\
$10^4$ & $10^4$ &  -0.8 & 1.0 &0.5 & 3.0  & 306. & 477. &1186. &114. &0.794   &0.153   \\
$10^4$ & $10^4$ &  -0.5 & 3.0 & -0.6 & -2.0  & 247. & 337. &1161.  &106. &2.31   &0.210   \\
$10^4$ & $10^4$ & 0.9 & 2.0 & -0.8 & 5.0  & 487.  & 698.  &251. &51.9
&29.3    &0.115   \\
$10^4$ & $10^4$ &  -0.6 & 1.0 &0.2 & 3.0  & 235.  & 378. &1498. &171.&\
1.33   &0.178   \\
$10^4$ & $10^4$ &  -0.1 & 3.0 & -0.9 & 6.0  & 343. & 402. &2462. &444.&\
2.47   &0.200   \\  
\end{tabular}
\end{table}


\begin{references}

\bibitem{lp:87} V.~M. Lipunov, and K.~A. Postnov, Soviet Astr.
{\bf 31}, 228 (1987).

\bibitem{lpp:87} V.~M. Lipunov, K.~A. Postnov, 
and M.~E. Prokhorov, Astron. Astrophys.
{\bf 176}, L1 (1987).

\bibitem{Hils:1990} D. Hils, P.~L. Bender, and R.~F. Webbink, ApJ {\bf 360},
75 (1990).

\bibitem{Pre} K. Danzmann et al., LISA Pre-Phase A Report,
Max-Planck-Institut fur Quantenoptik, Report MPQ 208, Garching,
Germany (1996). 

\bibitem{Peterseim_etal:1996} M. Peterseim, O. Jennrich, and K. Danzmann,
Class. Quant. Grav. {\bf 13}, 279 (1996).  

\bibitem{vecchio_rate} A. Vecchio, Class. Quant. Grav., in press. 

\bibitem{bbr} M.~C. Begelman, R.~D. Blandford, and M.~J. Rees,
Nature {\bf 287} 307 (1980).

\bibitem{haris} T. Apostolatos, C. Cutler, G.~J. Sussman, and  K.~S. Thorne,
Phys. Rev. D {\bf 49} 6274 (1994).

\bibitem{alberto_inprogress} A. Vecchio, in progress.

\bibitem{alberto_curt} A. Vecchio and C. Cutler, in progress.

\bibitem{cutler_flanagan} C. Cutler, and E.~E. Flanagan, Phys. Rev. D {\bf 49}  2658 (1994).

\bibitem{WZ} L.~A. Wainstein and V.~D. Zubakov,{\it Extraction of
Signals from Noise} (Dover Publications, Inc., New York, 1962).  

\bibitem{faller_etal} J.~E. Faller, P.~L. Bender, J.~L. Hall, and
M.~A. Vincent, Proc. Colloquium ``Kilometric Optical Arrays in Space,''
Cargese, 23-25, Oct. 1984; ESA SP-226, April 1985.

\bibitem{foot2} Since the laser beam will be $\sim 20$ km wide when it reaches
the opposite spacecraft, it is completely impractical to use mirrors 
to bounce the light back and forth, as one does with the 
ground-based detectors. 
Label the vertices of the triangle A,B,C. The lasers at B and C are
phase-locked to the incoming beams from A, so that the satellites at 
B and C effectively function as amplifying mirrors.

\bibitem{300years} K.S. Thorne, in {\it 300 Years of Gravitation},
            ed. S.W. Hawking and W. Israel (Cambridge University
            Press, Cambridge, 1987), pp. 330-458.

\bibitem{giampieri} Giampieri, G., Mon. Not. R. Astron. Soc., in press.

\bibitem{pbender_private}P.~L. Bender, private communication.

\bibitem{bh}P.~L. Bender and D. Hils, private communication.  We note
that our definition of the noise spectral density follows the
convention of~\cite{300years}.

\bibitem{bh}P.~L. Bender and D. Hils, Class. and Quant. Grav., in press.

\bibitem{webbink_1984}R.~F. Webbink, Ap.~J. {\bf 277}, 355 (1984).

\bibitem{draza}D. Markovi\'{c}, Phys. Rev. D. {\bf 48}, 4738 (1993).

\bibitem{3min} C. Cutler et al., Phys. Rev. Lett. {\bf 70}, 2984 (1993).

\bibitem{schilling} R. Schilling, Class. Quant. Grav., in press. 

\end{references}
\end{document}